\documentclass[runningheads,a4paper]{llncs}

\usepackage{amssymb}
\usepackage{graphicx}
\usepackage{graphics}
\usepackage{algorithm}
\usepackage{algorithmic}
\usepackage{url}
\urldef{\mailsa}\path|ioannis.paparrizos@epfl.ch|
\urldef{\mailsb}\path|anna.kramer, leonie.kunz, christine.reiss, nicole.sator,|
\urldef{\mailsc}\path|erika.siebert-cole, peter.strasser, lncs}@springer.com|    
\newcommand{\keywords}[1]{\par\addvspace\baselineskip
\noindent\keywordname\enspace\ignorespaces#1}

\begin{document}

\mainmatter  % start of an individual contribution

\title{A tight bound on the worst-case number of comparisons for Floyd's heap construction algorithm}

\titlerunning{A tight bound for Floyd's heap construction algorithm}

\author{Ioannis K. Paparrizos}
%
%\authorrunning{A tight bound for Floyd's heap construction algorithm}

\institute{School of Computer and Communication Sciences\\
\`Ecole Polytechnique F\`ed\`erale de Lausanne\\
\mailsa\\
}


\toctitle{Lecture Notes in Computer Science}
\tocauthor{Authors' Instructions}
\maketitle

\begin{abstract}
In this paper a tight bound on the worst-case number of comparisons for Floyd's well known heap construction algorithm, is derived. It is shown that at most $2n - 2\mu(n) - \sigma(n)$ comparisons are executed in the worst case, where $\mu(n)$ is the number of ones and $\sigma(n)$ is the number of zeros after the last one in the binary representation of the number of keys $n$.
\end{abstract}

\section{Introduction}
Floyd's heap construction algorithm \cite{paper4} proposed in 1964 as an improvement of the construction phase of the classical heapsort algorithm introduced earlier that year by Williams J.W.J \cite{paper8} in order to develop an efficient in-place general sorting algorithm. The importance of heaps in representing priority queues and speeding up an amazing variety of algorithms is well documented in the literature. Moreover, the classical heapsort algorithm and, hence, Floyd's heap construction algorithm as part of it, is contained and analyzed in each textbook discussing algorithm analysis, see \cite{paper1} and \cite{paper2} for example.

Floyd's algorithm is optimal as long as complexity is expressed in terms of sets of functions described via the asymptotic symbols $O$, $\Theta$ and $\Omega$. Indeed, its linear complexity $\Theta(n)$, both in the worst and best case, cannot be improved as each object must be examined at least once. However, it is an established tradition to analyze algorithms solving comparison based problem by counting mainly comparisons, see for example Knuth \cite{paper5} who states that the theoretical study of comparison counting gives us a good deal of useful insight into the nature of sorting processes.

Despite the overwhelming attention received by the computer community in the more than 45 years of its life, a tight bound on the worst-case number of comparisons holding for all values of $n$, is, to our knowledge, still unknown. Kruskal et al. \cite{paper6} showed that $2n - 2 \lceil \log(n+1) \rceil$
 is a tight bound on the worst-case number of comparisons, if $n = 2^k-1$, where $k$ is a positive integer, and, to our knowledge, this is the only value of $n$ for which a tight upper bound has been reported in the literature. 

Schaffer \cite{paper7} showed that $n -  \lceil \log(n+1) \rceil + \lambda(n)$, where $\lambda(n)$ is the number of zeros in the binary representation of $n$, is the sum of heights of sub-trees rooted at internal nodes of a complete binary tree, see also \cite{paper3} for an interesting geometric approach to the same problem. Using this result we show that $2n - 2\mu(n) - \sigma(n)$ is a tight bound on the worst-case number of comparisons for Floyd's heap construction algorithm. Here, $\mu(n)$ is the number of ones and $\sigma(n)$ is the number of zeros after the last (right most) one in the binary representation of the number of keys $n$.

\section{Floyd's heap construction algorithm}
A $maximum \hspace{2 pt} heap$ is an array $H$ the elements of which satisfy the property: 
\begin{equation}
H ( \lfloor i/2 \rfloor ) \geq H(i),  i = 2, 3, ... , n. 
\end{equation}
Relation (1) will be referred to as the heap property. A $minimum \hspace{2 pt} heap$ is similarly defined; just reverse the inequality sign in (1) from $\geq$ to $\leq$. When we simply say a heap we will always mean a maximum heap. A nice property of heaps is that they can be represented by a $complete \hspace{2 pt} binary \hspace{2 pt} tree$. Recall that a complete binary tree is a binary tree in which the root lies in level zero and all the levels except the last one contain the maximum possible number of nodes. In addition, the nodes at the last level are positioned as far to the left as possible. If $n = 2^k - 1$, the last level  $\lfloor log n \rfloor = k - 1$ contains the maximum possible number $2^k - 1$ of nodes. In this case the complete binary tree is called $perfect$. The $distinguished \hspace{2 pt} path$,  introduced in \cite{paper5}, of a complete binary tree that connects the root node $1$ with the last leaf node $n$, will play an important role in deriving our results. It is well known, see for example \cite{paper5}, that the nodes of the distinguished path correspond to the digits of the binary expression of $n$. \emph{Figure 1} illustrates a complete binary tree, its distinguished path and the corresponding binary expression of $n$. In terms of binary trees the heap property is stated as follows:\\

\emph{The value of a child is smaller than or equal to the value of its parent.}\\

\begin{figure}
\centering
\includegraphics[height=7cm,width=12cm]{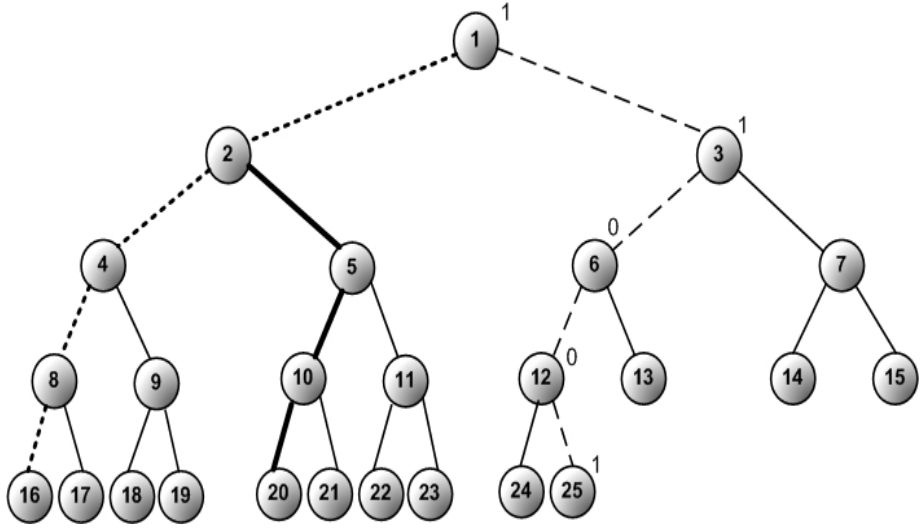}
\caption{A complete binary tree, its distinguished path (dashed edges), a special path (thick edges) and the leftmost path (dotted edges). The numbers by the nodes of the distinguished path are the digits of the binary expression (11001) of $n=25$.}
\label{fig:figure12}
\end{figure}

It is easily verified that the value of the root is the largest value. Also, each sub-tree $T_j$ of the complete binary tree representing a heap, is also a heap and, hence, the value $H(j)$ is the largest value among those that correspond to the nodes of $T_j$.  A sub-array $H(i:n)$ for which the heap property is satisfied by each node $j$ the parent of which is an element of $H(i:n)$, is also called a heap. Here the expression $j:n$ denotes the sequence of indices $j, j+1, j+2, ... , n.$

An \emph{almost heap} is a sub-array $H(i:n)$ all nodes of which satisfy the heap property except possibly node $i$. If key $H(i)$ violates the heap property, then $H(i) < \max\{H(2i), H(2i+1)\}.$

The main procedure of Floyd's heap construction algorithm, called in this paper heapdown, works as follows. It is applied to an almost heap $H(j:n)$ and converts it into a heap. In particular, if $m = H(j)$ satisfies the heap property $H(j) \geq \max\{H(2j), H(2j+1)\}$ and, hence, $H(j:n)$ is a heap, the algorithm does nothing. Otherwise, it swaps key $m = H(j)$ with the maximum child key $H(j_{max})$. Then, it considers the child $j_{max}$ which currently contains key $m$, and repeats the procedure until the heap property is restored. \emph{Algorithm 1} shows a formal description of the algorithm.

% put algorithm here
\begin{algorithm}
\caption{HEAPDOWN(H(i...n))}
\label{alg1}
\begin{algorithmic}
\WHILE{$2i + 1 \leq n$ }
\STATE $k = 2i$
\IF{$H(k) < H(k + 1)$}
\STATE $k = k+1$
\ENDIF
\IF{$H(i) < H(k)$}
\STATE $swap(H(i),H(k))$
\STATE $i = k$
\ELSE
\RETURN $H(i...n)$
\ENDIF
\ENDWHILE
\IF{$2i = n$ and $H(i) < H(n)$}
\STATE $swap(H(i),H(n))$
\RETURN $H(i...n)$
\ENDIF
\end{algorithmic}
\end{algorithm}

Floyd's heap construction algorithm, called Floyd - buildheap procedure in this paper, applies procedure heapdown to the sequence of almost heaps 
\begin{equation}
H ( \lfloor n/2 \rfloor  : n), H ( \lfloor n/2 \rfloor  - 1 : n), ... , H(1:n). 
\end{equation}
As the sub-array $H( \lfloor n/2 \rfloor +1 : n)$, consists of leafs and, therefore, it is a heap and procedure heapdown converts an almost heap to a heap, the correctness of procedure Floyd - buildheap is easily shown.

When procedure heapdown is applied to the almost heap $H(j:n)$ key $m = H(j)$ moves down one level per iteration. In general, two comparisons are executed per level, one comparison to find the maximum child and one to determine whether key $m$ should be interchanged with the maximum child key. However, there is a case in which just one comparison is executed. This happens when key $m$ is positioned at node  $ \lfloor n/2 \rfloor$  and $n$ is even. Then, internal node $n/2$ has just one child, the last node $n$, and therefore no comparison is needed to find the maximum child. We will see in the next section, when we will investigate the worst case of procedure Floyd-buildheap, that this situation happens quite often, if $n$ is even. Procedure FLOYD - BUILDHEAP describes formally Floyd's algorithm.

% put an algorithm here

%\begin{pseudocode}{FLOYD-BUILDFEAP}{H}
%\FOR i \GETS $\lfloor n/2 \rfloor$ \DOWNTO 1 with step $-1$ \DO
%heapdown(H(i...n))\\
%\RETURN{H}
%\end{pseudocode}

\begin{algorithm}
\caption{FLOYD-BUILDFEAP(H)}
\label{alg2}
\begin{algorithmic}
\FOR{$i = \lfloor n/2 \rfloor$ to $1$ step -1}
\STATE heapdown(H(i...n))
\ENDFOR
\RETURN $H$
\end{algorithmic}
\end{algorithm}

\section{A tight bound on the worst-case number of comparisons}

It is well known that the number of interchanges performed by Floyd's heap construction algorithm is bounded above by the sum $t(n)$ of heights of sub-trees rooted at the internal nodes of a complete binary tree. Schaffer \cite{paper7} showed that:
\begin{equation}
t(n) = n -  \lceil \log (n+1) \rceil + \lambda(n), 
\end{equation}
where $\lambda(n)$ is the number of zeros in the binary representation of $n$. For the sake of completeness of the presentation we provide a short proof based on the geometric idea described in \cite{paper3}. We associate a \emph{special path} with each internal node of the binary tree. The special path connects a node, say $j$, with a leaf of the subtree $T_j$ rooted at node $j$. The first edge of the special path is a \emph{right edge} and all the remaining edges are \emph{left edges}, see \emph{Picture 1}. In particular, the nodes of the special path are $j, 2j + 1, 2^2j+2, 2^3j+2^2, …, 2^kj+2^{k-1}$. Observe now that the edges of all special paths cover all the edges of the binary tree exactly ones except the  $\lfloor logn \rfloor$  edges of the \emph{leftmost path}, see \emph{Picture 1}. As no two special paths contain a common edge, the number of edges of all special paths is $n - 1 -  \lfloor logn \rfloor$.
  
The lengths of special paths are closely related to the heights of the sub-trees. Recall that the \emph{length} of a path is the number of edges it contains. Denote by $sp(j)$ the special path corresponding to node $j$. If internal node $j$ does not belong in the distinguished path, then $length(sp(j)) = h(T_j)$. If internal node $j$ belongs in the distinguished path and the right edge $(j, 2j+1)$ is an edge of the distinguished path, then $length(sp(j)) = h(T_j)$. In that case the first edge of $sp(j)$ belongs in the distinguished path and the digit of the binary expression of $n$ corresponding to node $j$ is $1$. If internal node $j$ belongs in the distinguished path and the left edge $(j, 2j)$ is an edge of the distinguished path, then $h(T_j) = length(sp(j)) + 1$. In that case the first edge of $sp(j)$ does not belong to the distinguished path and the digit of the binary expression of $n$ corresponding to node $j$ is $0$. Summing up all heights of internal nodes we get 
\begin{equation}
t(n) = n -1 - \lfloor logn \rfloor  + \lambda(n) = n - \lceil log (n+1)\rceil  + \lambda(n).
\end{equation}
In computing, our tight bound on the worst-case number of comparisons, two cases must be considered, $n$ even and $n$ odd. We first take care of the case $n$ is odd. 

\subsubsection{Lemma 1.} Let $n$ be odd. Then the maximum number of comparisons executed by Floyd's heap construction algorithm is 
\begin{equation}
2t(n) = 2(n-  \lceil \log (n+1) \rceil + \lambda(n)).
\end{equation}
\subsubsection{Proof:} If $n$ is odd, each internal node has exactly two children and, hence, each key swap corresponds to two key comparisons. Therefore $2t(n)$ is an upper bound on the number of comparisons. 

We show now that this bound is tight. To this end we construct a special worst case array $H$. In particular $H$ satisfies the following properties 

\begin{enumerate}
\item The elements of $H$ are the $n$ distinct keys $1,2, ... ,n$.
\item The nodes in the distinguished path are assigned the  $ \lceil \log (n+1) \rceil $  largest keys. In particular, the nodes in levels $0, 1, 2, ... ,  \log(n)$  are assigned the keys  $n - \lceil \log (n+1) \rceil + 1, n -  \lceil \log (n+1) \rceil + 2, ... , n$ respectively.
\item If $j$ is a node not belonging in the distinguished path, sub-tree $T_j$ is a minimum heap.
\end{enumerate}

Apply now procedure Floyd-buildheap to the array $H$ described previously. When procedure heapdown is called on the almost heap $H(j:n)$ and $j$ is not a node of the distinguished path, key $m = H(j)$ will move all the way down to the bottom level of sub-tree $T_j$. This is so because key $m$ is the smallest among the keys corresponding to nodes of the sub-tree rooted at node $j$, see property 3. Also, two comparisons are executed per level. When procedure heapdown is applied to an almost heap $H(j : n)$, where $j$ is a node of the distinguished path, key $m = h(j)$ will follow the distinguished path all the way down to the bottom level taking the position of leaf node $n$, see property 2. Again, two comparisons are executed per level and, hence, the number 
$2n - 2 \lceil \log (n+1) \rceil + 2\lambda(n)$ 
is a tight bound on the worst-case number of comparisons. $\Box$      

Next lemma takes care of the case $n$ even.

\subsubsection{Lemma 2.} If $n$ is even the exact worst case number of comparisons for Floyd's heap construction algorithm is 
\begin{equation}
2(n -  \lceil \log (n+1) \rceil + \lambda(n)) - \sigma(n), 
\end{equation}
where $\sigma(n)$ is the number of zeros after the last one in the binary representation of $n$.
\subsubsection {Proof:} Let $(b_m b_{m-1} ... b_2 b_1 b_0)$ be the binary representation of $n$. Let also $b_k b_{k-1} ...\\ b_2 b_1 b_0$ be the last $k+1$ digits of the binary representation of $n$ such that $b_k = 1$ and $b_{k-1} = b_{k-2} = ... = b_1 = b_0 = 0$.

As $n$ is even $b_0 = 0$ and, hence, $k \geq 1$. Consider now an internal node of height $j \leq k$ lying at the distinguished path. It is easily verified, using inductively the well known property $\lfloor \lfloor n/2 \rfloor  / a \rfloor = \lfloor n / a^2 \rfloor $
of the floor function, that the index at that node is  $ \lfloor n / 2^j \rfloor $ . When procedure Floyd-buildheap calls procedure heapdown on the almost heap $H( \lfloor n / 2^j \rfloor  : n)$ key $m = H( \lfloor n / 2^j \rfloor )$ will move down the levels either following the distinguished path or moving to the right of it at some point. This is so because all the edges $(n,  \lfloor n / 2 \rfloor), ( \lfloor n / 2 \rfloor , \lfloor n / 2^2 \rfloor ), ... , ( \lfloor n / 2^{j-1} \rfloor , \lfloor n / 2^j \rfloor)$ of the distinguished path are left edges. In the former case at most $2j - 1$ comparisons are executed and this happens when key $H( \lfloor n / 2^j \rfloor)$ is placed either at the bottom level or at the level next to bottom. In the latter case at most $2(j - 1)$ comparisons are executed. Hence for each node of the distinguished path at height $j = 1, 2, ... , k$ the maximum number of comparisons is one less than 2 times the height of the sub-tree rooted at that node. For all the remaining internal nodes $i$ the maximum number of comparisons is $2h(T_i)$, where $h(T_i)$ is the height of the sub-tree rooted at node $i$. As the number of internal nodes of the distinguished path at heights $1, 2, ... , k$ is $\sigma(n)$,  the previous arguments show that the number
\begin{equation}
2(n -  \lceil \log (n+1) \rceil + \lambda(n)) - \sigma(n)
\end{equation}
is an upper bound on the number of comparisons for procedure Floyd-buildheap.

\begin{figure}
\centering
\includegraphics[height=8cm,width=12cm]{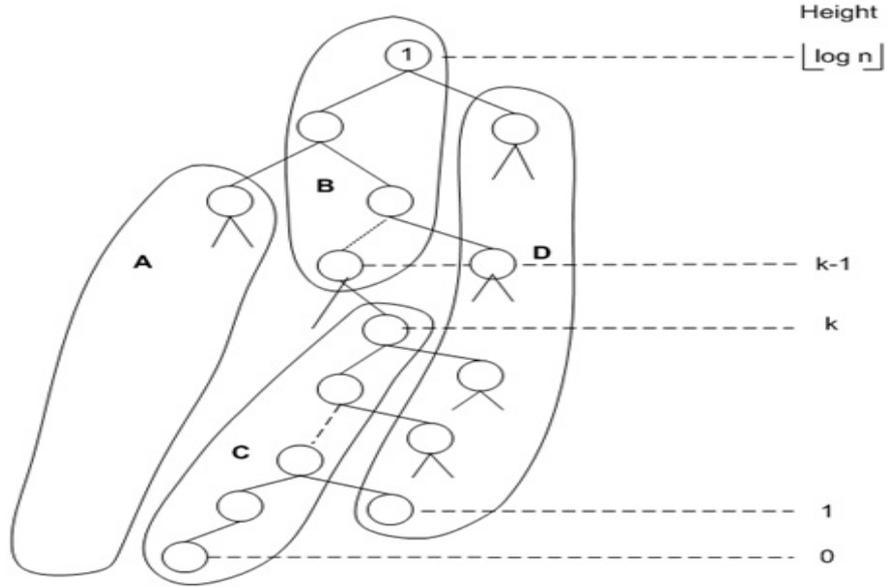}
\caption{Partition of the nodes of a complete binary tree into sets $A, B, C, D$.}
\label{fig:figure1}
\end{figure}

We describe now an array $H$ on which procedure Floyd-buildheap executes exactly $2(n -  \lceil \log (n+1) \rceil + \lambda(n)) - \sigma(n)$ comparisons, thus showing that this number is a tight upper bound for $n$ even. In order to describe the structure of the worst case example $H$ we partition the nodes of the complete binary tree into 4 sets $A, B, C, D$. Set $A$ contains all the nodes on the left side of the distinguished path. Set $D$ contains all nodes lying on the right side of the distinguished path. Set $C$ contains the nodes of the distinguished path of height $j = 0, 1, 2, ... , k$ and set $B$ contains all the remaining nodes of the distinguished path. \emph{Figure}~\ref{fig:figure1} illustrates a complete binary tree and the sets of nodes $A, B, C, D$.

The structure of array $H$ is described in the following properties:
\begin{enumerate}
\item The elements of $H$ are the $n$ distinct keys $1, 2, ... , n$. 
\item If  $i, j, k, l$ are nodes belonging to the sets $A, B, C, D$ respectively, then 
\begin{equation}
H(i) > H(j) > H(k )> H(l).
\end{equation}
\item The keys in the distinguished path that belong to the set $B$ are in increasing order from the top to the bottom. The keys in the distinguished path that belong to the set $C$ are in increasing order from the top to the bottom.
\item If $j$ is a node not belonging to the distinguished path, the sub-tree $T_j$ is a minimum heap.
\end{enumerate}

\begin{figure}
\centering
\includegraphics[height=8cm,width=12cm]{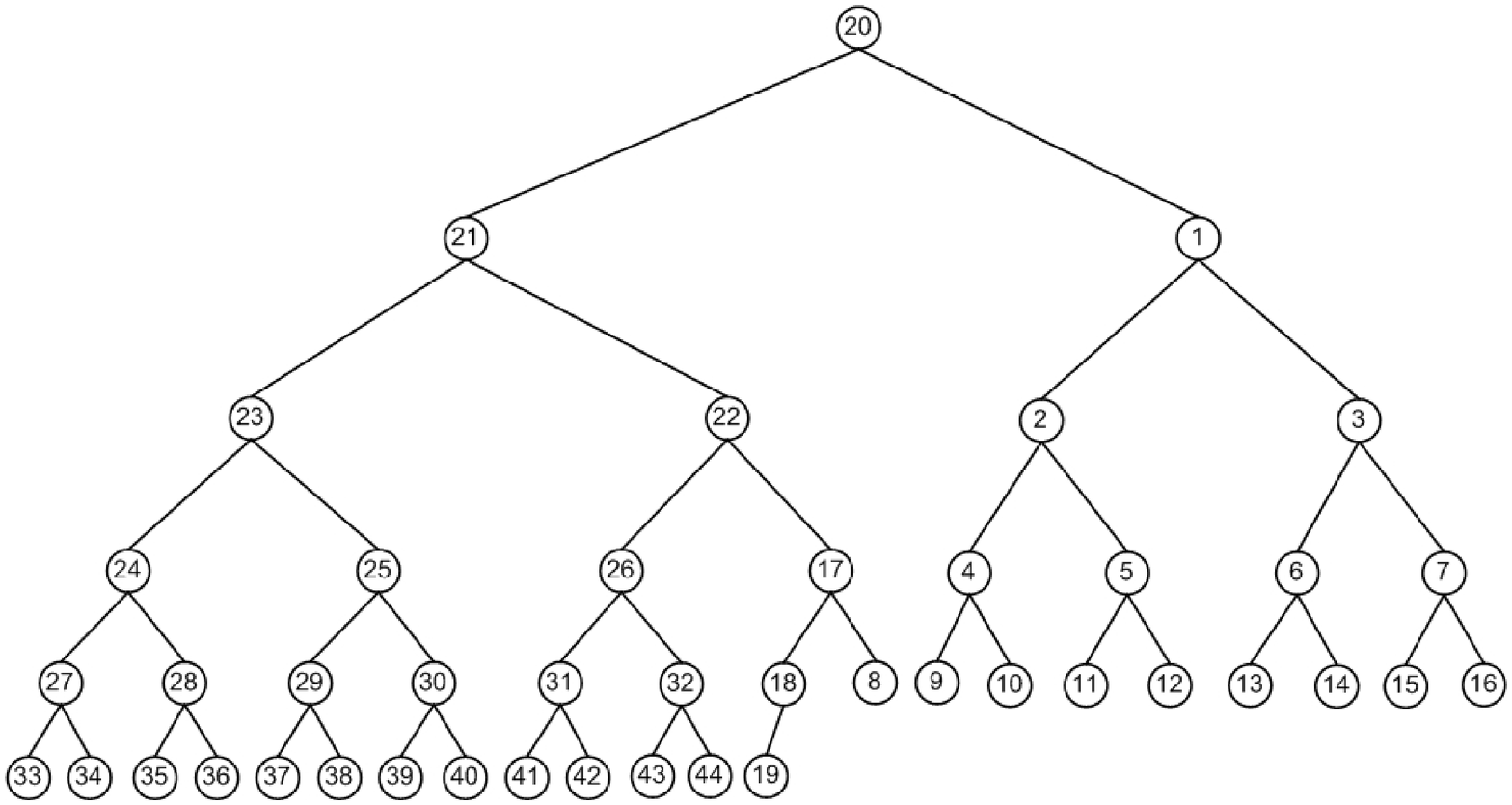}
\caption{A worst case complete binary tree for Lemma 2. It is $n = 44$, $k = 2$, $|A| = 23, |B| = 3, |C| = 3, |D| = 15$. The number inside node $j$ is the key $H(j)$.}
\label{fig:figure2}
\end{figure}

Although there are more than one way to assign the keys $1, 2, ..., n$ to the elements of $H$ so that properties 2), 3) and 4) are satisfied, an easy way to do that is as follows. Place the $|A|$ largest keys to the sub-trees on the left of the distinguished path so that each sub-tree is a minimum heap. The symbol $|A|$ denotes the number of elements of set $A$. Obviously $|A|$ is the number of nodes on the left of the distinguished path. Place in increasing order from top to bottom levels the next $|B|$ largest elements at the nodes of the distinguished path that belong to the set $B$. Also, place in increasing order from top to down levels the next $|C|$ largest elements at the nodes of the distinguished path that belong to the set $C$. Obviously 
$|B| + |C| = 1 +  \lfloor \log(n) \rfloor = \lceil \log(n+1) \rceil$. 
Finally, place the remaining $|D|$ smallest keys, i.e, the keys $1,2, ... ,|B|$ at the sub-trees right to the distinguished path so that each sub-tree is a minimum heap. \emph{Figure}~\ref{fig:figure2} illustrates such a worst case example for $n = 44$.

Apply now procedure Floyd-buildheap on the array $H$ described previously. Let
$H(j:n), j =  \lfloor n / 2 \rfloor, \lfloor n / 2 \rfloor -1, ... , 1$
be the almost heap on which procedure heapdown is applied to after it is called by procedure Floyd-buildheap. If $j$ is not a node at the distinguished path, key $H(j)$, because of property 4), will move all the way down to the bottom level of sub-tree $T_j$ and $2h(T_j)$ comparisons will be executed. If $j$ is a node of the distinguished path belonging to set $C$, key $H(j)$, because of properties 2), 3) and 4), will follow the distinguished path never making a right turn. In this case, key $H(j)$ will be placed at node $n$ executing $2h(T_j) - 1$ comparisons. Finally, if node $j$ belongs in set $B$, key $H(j)$ will definitely make a left turn before reaching the node  $\lfloor n / 2^k \rfloor$  of the distinguished path, see properties 2) and 3). Then it will move all the way down to bottom level executing 2 comparisons per level. Again $2h(T_j)$ comparisons are executed.

Summing up the comparisons for all the calls of procedure heapdown we see that the total number of comparisons is as stated in the Lemma. $\Box$ 
                     
Observe that the array $H$ described in the previous Lemma is not a minimum heap. In particular the sub-trees rooted at nodes of the distinguished path are not minimum heaps.

\subsubsection{Theorem 1.} The number $2n - 2\mu(n) - \sigma(n)$, where $\mu(n)$ is the number of ones and $\sigma(n)$ is the number of zeros after the last one in the binary representation of $n$, is a tight bound on the worst-case number of comparisons for Floyd's heap construction algorithm.

\subsubsection{Proof:} If $n$ is odd, then $b_0 = 1$ and, hence, $\sigma(n) = 0$. Combining Lemmas 1 and 2 we see that a tight bound on the worst-case number of comparisons is the number $2[n - \lceil \log (n+1) \rceil + \lambda(n)] - \sigma(n) = 2[n - (\lambda(n) + \mu(n)) + \lambda(n)] - \sigma(n) = 2n- 2\mu(n) - \sigma(n).$ $\Box$

\section{Conclusion}

Deriving worst case tight upper bound examples for an algorithm implies that the worst case complexity of the algorithm cannot be improved. We derived our worst case examples by the use of simple geometric ideas. As the binary trees and heaps are involved in many other algorithms for which worst case tight examples are not known, we hope that our results will contribute in solving those problems.

\end{document}